\documentclass[twocolumn,showpacs,preprintnumbers,amssymb]{revtex4}

\usepackage{graphicx}
\usepackage{epsfig}
\usepackage{dcolumn}
\usepackage{bm}
\def\gsim{\lower0.5ex\hbox{$\:\buildrel >\over\sim\:$}}
\def\lsim{\lower0.5ex\hbox{$\:\buildrel <\over\sim\:$}}

\newcommand{\be}{\begin{equation}}
\newcommand{\ee}{\end{equation}}
\newcommand{\bea}{\begin{eqnarray}}
\newcommand{\eea}{\end{eqnarray}}

\newcommand{\nbox}{{\,\lower0.9pt\vbox{\hrule \hbox{\vrule height 0.2 cm
\hskip 0.2 cm \vrule height 0.2 cm}\hrule}\,}}

\def\missET {{\not\!\! E_T}}

\begin{document}

\preprint{UCI-TR-2008-13}

\title{Collider Signals of Maximal Flavor Violation:\\ Same-Sign Leptons from Same-Sign Tops at the Tevatron}

\author{Shaouly Bar-Shalom$^{a,b}$}
\email{shaouly@physics.technion.ac.il}
\author{Arvind Rajaraman$^b$}
\email{arajaram@uci.edu}
\author{Daniel Whiteson$^b$}
\email{daniel@uci.edu}
\author{Felix Yu$^b$}
\email{felixy@uci.edu}
\affiliation{$^a$ Physics Department, Technion-Institute of Technology,
Haifa 32000, Israel\\
$^b$Department of Physics and Astronomy, University of California,
Irvine, CA 92697, USA}

\date{\today}

\begin{abstract}
 In models of maximal flavor violation (MxFV) suggested in \cite{ourMxFV}
 there is at least one new scalar $\Phi_{FV}$ which couples to the quarks via
 $\Phi_{FV} q_i q_j \propto \xi_{ij}$ where $\xi_{i3},\xi_{3i} \sim V_{tb}$
for $i = 1,2$ and $\xi_{33} \sim V_{td}$ and $V$ is the CKM matrix.
In this article, we explore the potential phenomenological implications
of MxFV for collider experiments. We study MxFV signals of same-sign
leptons from same-sign top-quark pair production at the Tevatron and
at the LHC. We show that
the current Tevatron dataset has strong sensitivity to this signature, for which there are no current limits. For example, if $m_{\Phi_{FV}} \sim 200$ GeV and the MxFV coupling $\xi$ has a natural value of $\sim 1$, we expect $\sim 12$ MxFV events to survive a selection requiring a pair of same-sign leptons, a tagged $b$-jet and missing transverse energy, over a background of approximately 4-5 events. 
\end{abstract}


\maketitle

If there is New Physics (NP) around the TeV scale, as
suggested
by the hierarchy problem and the existence of dark matter, then
flavor violating (FV) processes can in  principle
 occur at large rates, but such processes are not observed. This implies that there is some structure
to the new physics couplings. 
One such structure is 
the minimal FV (MFV) ansatz, which states that the NP
is ``aligned'' with the SM, such that
all FV transitions are governed by the nearly diagonal CKM matrix $V$.
The MFV ansatz therefore imposes the couplings of any new scalar to
a pair of top+light quark to satisfy
$\xi_{3i}, \xi_{i3} (\sim V_{td}) \ll \xi_{33} (\sim V_{tb})$ for $i=1,2$.

In a previous paper~\cite{ourMxFV}, two of the authors have
presented a new class of scalar-mediated MxFV models
which (as suggested by their name) maximally depart from the MFV ansatz
in the sense that $\xi_{31},\xi_{32} \sim {\cal O}(1) \gg \xi_{33}$,
and still satisfy all constraints from flavor physics
even with a relatively light scalar with a mass of ${\cal O}(m_W)$.

In particular, let $\Phi_{FV} \equiv (\eta^+,\eta^0)$ be a
 new scalar doublet that mediates MxFV through \cite{ourMxFV}:
\begin{eqnarray}
{\cal L}_{FV} &=& \xi_{ij}  \bar Q_{iL} \tilde\Phi_{FV} 
u_{jR} + h.c. ~,\label{LFV}
\end{eqnarray}
\noindent where $\xi$ is a 3x3 matrix in flavor space.
In~\cite{ourMxFV}, it was shown that some substructures of the
MxFV texture:
\begin{eqnarray}
\xi \equiv \begin{pmatrix} {
0 & 0 & \xi_{13} \cr
0 & 0 & \xi_{23} \cr
\xi_{31} & \xi_{32} & 0 } \end{pmatrix}
\label{texture}~,
\end{eqnarray}
can potentially avoid constraints from low-energy
flavor data such as meson mixings and K-decays.
In particular, it was shown
in \cite{ourMxFV} that there are no constraints if only one MxFV coupling is
non-zero
(e.g., the case $\xi_{31} \sim {\cal O}(1) \gg \xi_{13},\xi_{32},\xi_{23}$
is not ruled out regardless of $m_{\eta^0}$ and $m_{\eta^+}$),
 and that the MxFV$_1$ models (defined as models with
$\xi_{31},\xi_{13} \gg \xi_{32},\xi_{23}$)
are not ruled out even with
$\xi_{31},\xi_{13} \sim {\cal O}(1)$, as long as $m_{\eta^+} \gsim
600$ GeV (regardless of $m_{\eta^0}$).\\

A list of possible collider signals of MxFV models was given in \cite{ourMxFV}. In this paper we study in detail one possible signal; same-sign charged leptons from same-sign top quark pair production.
For definiteness, in what follows we will study the case of MxFV$_1$ models
(defined above) under the assumptions that 
$\xi_{ij}$ are real and that
$\xi \equiv \xi_{31} = \xi_{13}$. As was shown in~\cite{ourMxFV}, in this case the only relevant
(sizable) flavor changing couplings are:

\begin{eqnarray}
\Gamma_{\eta^0 \bar t u} = \Gamma_{\eta^0 \bar u t} = - i \xi ~,~
\Gamma_{\eta^+ \bar t d} = \Gamma_{\eta^+ \bar u b} = \frac{i}{2} \xi \left(1 - \gamma_5\right) \label{FR1}.
\end{eqnarray}

A particularly interesting limit which we will study in this paper
is when the charged scalar $\eta^+$ is
too heavy to be accessible at Tevatron and LHC energies, and
decouples (the sensitivity of the LHC and the Tevatron to $\eta^+$
will be discussed in a separate paper).  If the neutral scalar is
light ($m_{\eta^0} \ll m_{\eta^+}$) it can still be probed at
colliders. Note that in this case, there are essentially no constraints from
low energy data.

The neutral scalar decays half the time to $t+\bar
u$ and half the time to $ \bar t + u$. This leads to a striking
signal, because we can have
{\it production of same-sign top-quark pairs in association with
light-quark jets} through the processes:

\begin{eqnarray}
 && ug \to t \eta^0 \to t t \bar u + h.c.~, \label{ttu1}
\\
 && u \bar u \to \eta^0 \eta^0 \to t t \bar u \bar u + h.c.~,
 \label{ttuu1}
\\
 && u u \to t t +h.c. ~, 
 \label{tt1}
 \label{ttu2}
\end{eqnarray}

\noindent where the last process comes from t-channel $\eta^0$ { exchange}.$^{[1]}$\footnotetext[1]{The
$tt \bar u$ final state also receives an additional (sub-leading) contribution from the pure $2 \to 3$
t-channel $\eta^0$-exchange process $ug \to t t \bar u$ which is included in our analysis.} 
As we will see below, this same-sign top pair production signature will
be an unambiguous signal of MxFV; indeed, there is no such
process in the SM or any extension to the SM with natural flavor conservation like the MSSM.
Note that $t\bar{t}$
production (with additional jets) from MxFV $\eta^0$ decays is less sensitive
to the new physics because the SM background is larger.

Another important cross-check is that there is
no s-channel resonance of the new MxFV scalar, see \cite{ourMxFV}. In particular,
since only $\eta^+$ can be produced in resonance, the case of a decoupling
$\eta^+$ leads to a very distinct
Higgs
phenomenology as one would observe a new scalar (i.e., $\eta^0$) produced
in pairs or
in association with a top-quark, but, contrary to the usual expectations,
this new scalar would not be observed on resonance.

We now consider  the production of same-sign top-quark pairs at
the Tevatron through the processes mentioned above.
We define the inclusive reaction:
\begin{eqnarray}
p \bar p \to t t + nj +X  \label{ttinc}~,
\end{eqnarray}
where $tt$
stands for both the $tt$ and $\bar t \bar t$ production channels
(as we will be interested in same-sign leptons signals either 
positively or negatively charged)
and $n$ is the number of light-quark jets $j$, each with transverse energy $E_T>15$ GeV.
Note that, for $m_{\eta^0} > m_t$, $\hat\sigma(ug \rightarrow t \eta^0
\rightarrow tt \bar u) \propto \xi^2$ while $\hat\sigma(uu \rightarrow  tt)$
and $\hat\sigma(u\bar{u} \rightarrow  \eta^0 \eta^0 \rightarrow
tt \bar{u}\bar{u}) \propto \xi^4$. Thus,
for $\xi <1$, $\sigma(p \bar p \to t t + nj +X)$ is dominated by
the $t \eta^0$ production channel.

When both top quarks decay leptonically
($t\rightarrow Wb \rightarrow l\nu b$), these processes have a
striking low-background signature of two same-sign leptons,
missing energy, and two $b$-jets ($\ell^\pm \ell^\pm \missET bb$)
accompanied by
$n$ hard jets. Though CDF has examined its inclusive same-sign lepton dataset in small datasets\cite{cdfss}, there has not been an experimenal study of the $\ell^\pm \ell^\pm \missET bb$ final state in which many of the same-sign contributions are supressed by the requirement of a $b$-tag or missing transverse energy. In what follows we describe an event selection to isolate these same-sign lepton signatures, calculate the expected number of such
events in the Tevatron data, estimate the contributions from background
sources, and determine the sensitivity of a single Tevatron detector as a function of $\xi$ and
$m_{\eta^0}$.\\

\noindent {\it Event Reconstruction and Selection}

To isolate the same-sign top quarks signal we define the $l^\pm l^\pm b \missET$ signature by requiring:

\begin{itemize}
\item Two same-sign reconstructed leptons (electrons or muons), each
with $p_T > 20$ GeV/$c$.
\item At least one secondary-vertex tag ($b$-tag)~\cite{secvtx}.
\item At least 20 GeV of missing transverse energy, $\missET$.$^{[2]}$\footnotetext[2]{Missing transverse energy,
    $\missET$, is defined as the
magnitude of the vector, $ -\sum_i E_T^i \vec n_i$, where $E_T^i$ are the
magnitudes of transverse energy contained in each calorimeter tower $i$,
and $\vec n_i$ is the unit vector from the interaction vertex to the tower
in the transverse ($x,y$) plane. $\missET$ is further corrected for 
reconstructed jets and muons.}
\end{itemize}

\noindent{\it Expected Yield and Backgrounds}

To calculate the number of $tt$ and $\bar t \bar t$ events we
expect at a single Tevatron experiment, we generate events for each of the three same-sign
processes in (\ref{ttu1})-(\ref{tt1}) using Calchep\cite{calchep} and shower them using
{\sc pythia}~\cite{pythia}. Detector resolution and acceptance are modeled using a parametric detector simulation written to approximately describe the performance of a CDFII-like detector, including lepton acceptance and charge identification, missing energy resolution, misidentification of leptons from jets, $b$-tagging efficiency as well as mistagging of light-quark jets as $b$-jets.  The performance of our parametric detector simulation was compared to several published CDFII results and found to agree to within 25\%.  Table~\ref{tab:sigcdf} shows the number of expected events in 2 fb$^{-1}$ of data.

\begin{table}[h]
\caption{ Production cross-sections $\sigma(tt)$, $\sigma(tt\bar{u})$ and
$\sigma(tt\bar{u}\bar{u})$ for each of the three same-sign top quark processes
in (\ref{ttu1})-(\ref{ttu2}), for $\xi=1$ and various $\eta^0$ masses. Also
given are the acceptance ($\epsilon$) of the event selection described in the
text and expected number ($N$) of $l^\pm l^\pm b \missET$ events in 2 fb$^{-1}$ of data. The
uncertainty on the
cross-sections is estimated to be 10\%, mainly due to the choice of the
renormalization scale, the choice of PDF's and the numerical integration.}
\begin{ruledtabular}
\begin{tabular}{llcccccc}
&$M_{\eta^0}$ [GeV/c$^2$] & 180 & 190 & 200 & 225 & 250 & 300 \\ \hline
&$\sigma$ [pb] & 0.50 & 0.45 & 0.41 & 0.33 & 0.27 & 0.19 \\
$tt$&$\epsilon$ [\%] & 0.5 & 0.5 & 0.5 & 0.5 & 0.5 & 0.5 \\
&$N$ &  4.8 & 4.4 & 4.1 & 3.3 & 2.6 & 1.8 \\
\hline
&$\sigma$ [pb] & 0.54 & 0.50 & 0.42 & 0.28 & 0.22 & 0.10 \\
$tt\bar{u}$&$\epsilon$ [\%]& 0.5 & 0.5 & 0.5 & 0.5 & 0.5 & 0.5 \\
&$N$ &  5.3 & 4.9 & 4.3 & 3.0 & 2.4 & 1.1 \\
\hline
&$\sigma$ [pb] & 0.68 & 0.45 & 0.38 & 0.17 & 0.06 & 0.02 \\
$tt\bar{u}\bar{u}$&$\epsilon$ [\%] & 0.5 & 0.5 & 0.5 & 0.5 & 0.5&0.5 \\
&$N$ & 6.4 & 4.7 & 4.1 & 1.8 & 0.7 & 0.2 \\
\hline
\multicolumn{2}{l}{Total $N(l^\pm l^\pm b \missET$)} & 16.5 & 14.0 & 12.5 & 8.1 & 5.7 & 3.1 \\
\end{tabular}
\end{ruledtabular}
\label{tab:sigcdf}
\end{table}

Major backgrounds to the $l^\pm l^\pm b \missET$ signature come from:

\begin{itemize}
\item $Z+$jets $\rightarrow e^+e^-$+jets, in which the $e^+$ or
$e^-$ emits a hard photon which later converts asymmetrically in the detector, giving a
same-sign lepton pair.
\item $W+$jets, where one jet is misidentified as a lepton, typically an electron
\item $t\bar t$ events where $t \bar t \to bl\nu bjj$ and a second lepton
comes from semi-leptonic decays of one of the $b$ quarks, or
$t \bar t \to be^+\nu b e^-\nu$ with a same-sign $ee$ pair arising
from a trident.
\end{itemize}

\begin{figure}[htb]
\includegraphics[width=7cm]{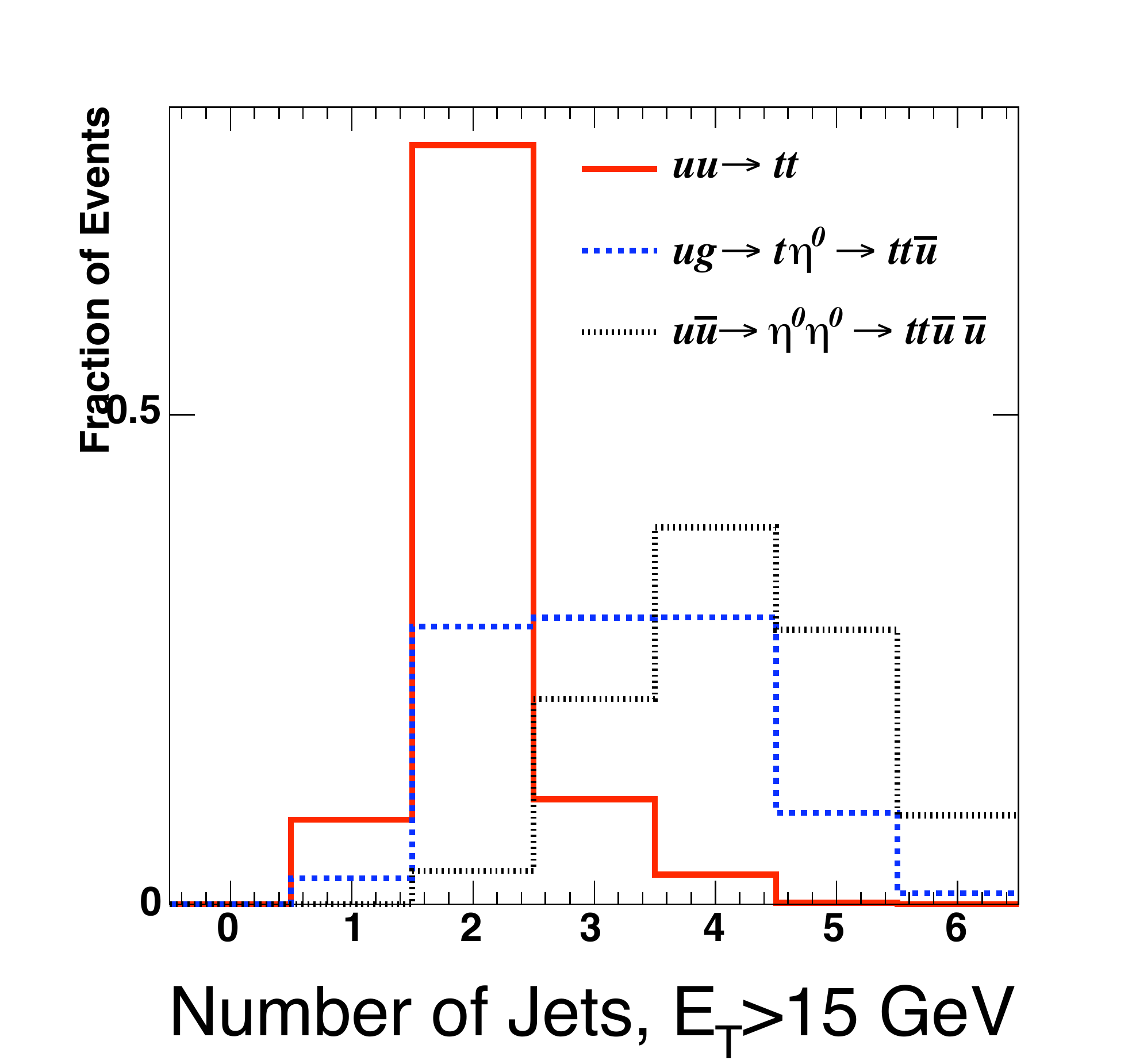}
\includegraphics[width=7cm]{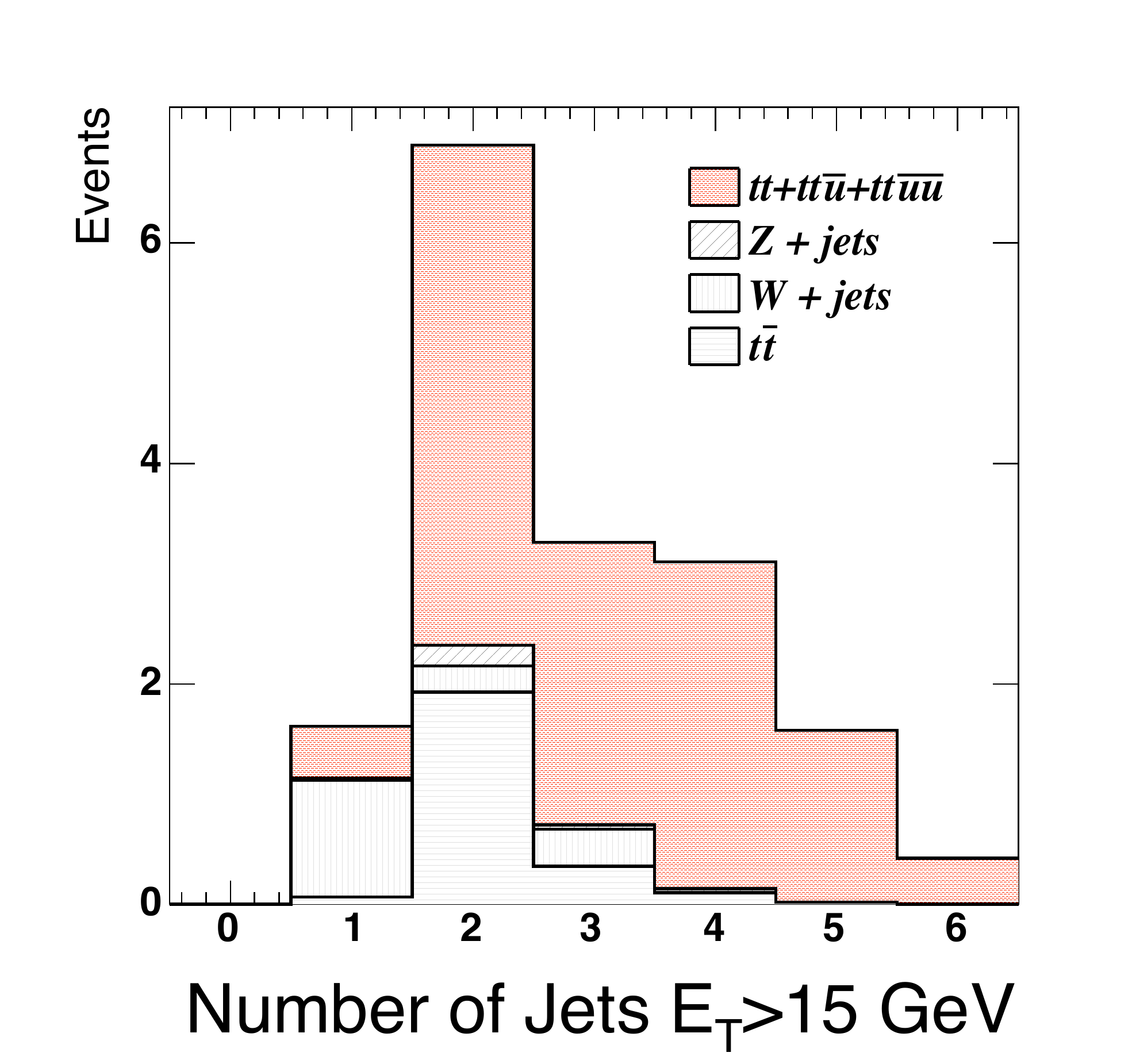}
\caption{Top: distribution in reconstructed jets with
$E_T>15$ GeV  for the signal process (each process is shown with unit area) 
with $m_{\eta^0}=200$ GeV/c$^2$
and after requiring same-sign leptons, 20 GeV of $\missET$ and at least one
$b$-tagged jet. Bottom: expected number
of reconstructed jets in 2 fb$^{-1}$ of data, for the signal processes with $\xi=1$ and
backgrounds.}
\label{fig:shapes}
\end{figure}

Backgrounds from diboson production $WW, WZ, ZZ, W\gamma$ and  $Z\gamma$ that
produce real same-sign leptons are found to be insignificant due to the
requirement of a $b$-tag.

Backgrounds from $Z$+jets processes are estimated
using {\sc alpgen}~\cite{alpgen} matched with {\sc pythia} for the
showering.   Kinematics of fake-lepton
backgrounds are described using {\sc alpgen} $W$+jet events for the hard
process with showering and matching as with $Z$+jets.  The
$t\overline{t}$ backgrounds are estimated using events
generated in {\sc pythia} at $m_t = 171$ GeV/c$^2$. As with the signal case,
 the efficiency of the selection after event reconstruction is described 
using a parametric detector simulation.  
Table~\ref{tab:bgcdf} shows the number of expected background
events in 2 fb$^{-1}$ of data.

\begin{table}[h]
\caption{Expected number of background events for the
$l^\pm l^\pm b \missET$ signature in 2 fb$^{-1}$ of data including
three categories of lepton flavors ($ee$,$\mu\mu$,$e\mu$)
, and the total background. Systematic uncertainties in the background are
dominated by uncertainties in the normalization corrections.}
\begin{ruledtabular}
\begin{tabular}{lr}
Source & $N(l^\pm l^\pm b \missET)$\\ \hline
$Z\gamma,W\gamma,WW,WZ,WW$ & $<0.2$\\
$Z$+jets & $0.3\pm0.2$\\
$W$+jets & $1.6\pm1.0$ \\
$t\overline{t}$ & $2.5\pm1.1$\\
\hline
Total & $4.4\pm1.5$ \\
\end{tabular}
\end{ruledtabular}
\label{tab:bgcdf}
\end{table}

\noindent{\it Sensitivity to $\xi$}

From the experimental data, one could measure directly the value of the MxFV
coupling $\xi$, which is directly proportional to
$\sigma(p \bar p \to t t +nj +X)$
at a specific $m_{\eta^0}$. The simplest method would be to transform
the number of observed events over the expected background into a
measurement of $\sigma(p \bar p \to t t +nj +X)$ and therefore of $\xi$.
To improve sensitivity, we simultaneously fit for the number of signal
and background events in the data by exploiting the difference between the
number of expected jets in signal and background events, see
Fig.~\ref{fig:shapes}; the fitted number of signal events can be
transformed into a fitted value for $\xi$.

\begin{figure}[htb]
\includegraphics[width=7.5cm]{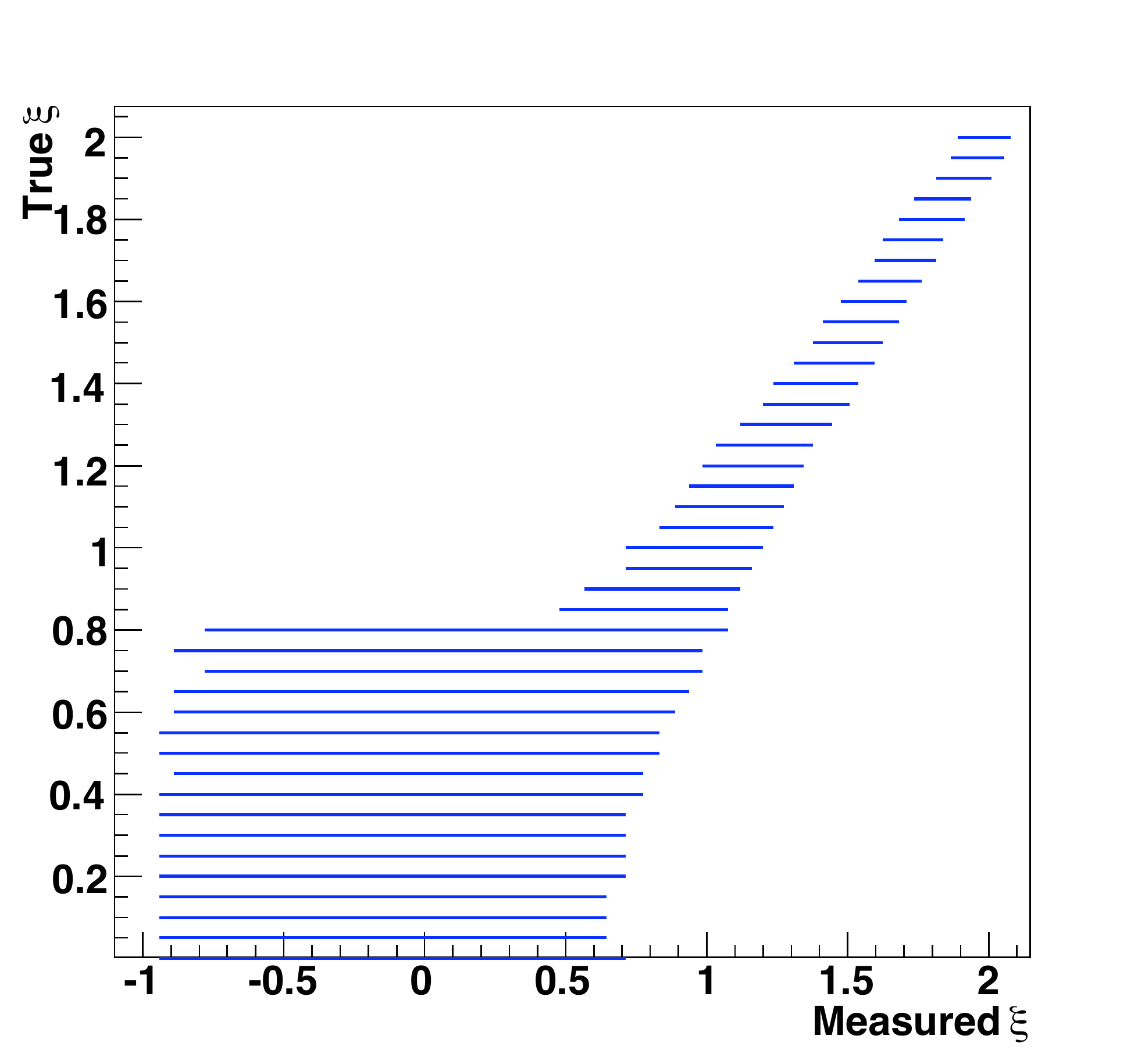}\\
\caption{Horizontal bands in fitted (measured) $\xi$ which include 95\% of the
results of Monte Carlo experiments, for varying values of true $\xi$,
with $m_{\eta^0}=200$ GeV/c$^2$,  following the prescription in~\cite{feldcous}.
A 95\% CL band in true $\xi$ for a given fit $\xi$ is a vertical band at the
measured value (see text).}
\label{fig:fc}
\end{figure}

\begin{figure}[htb!]
\includegraphics[width=7cm]{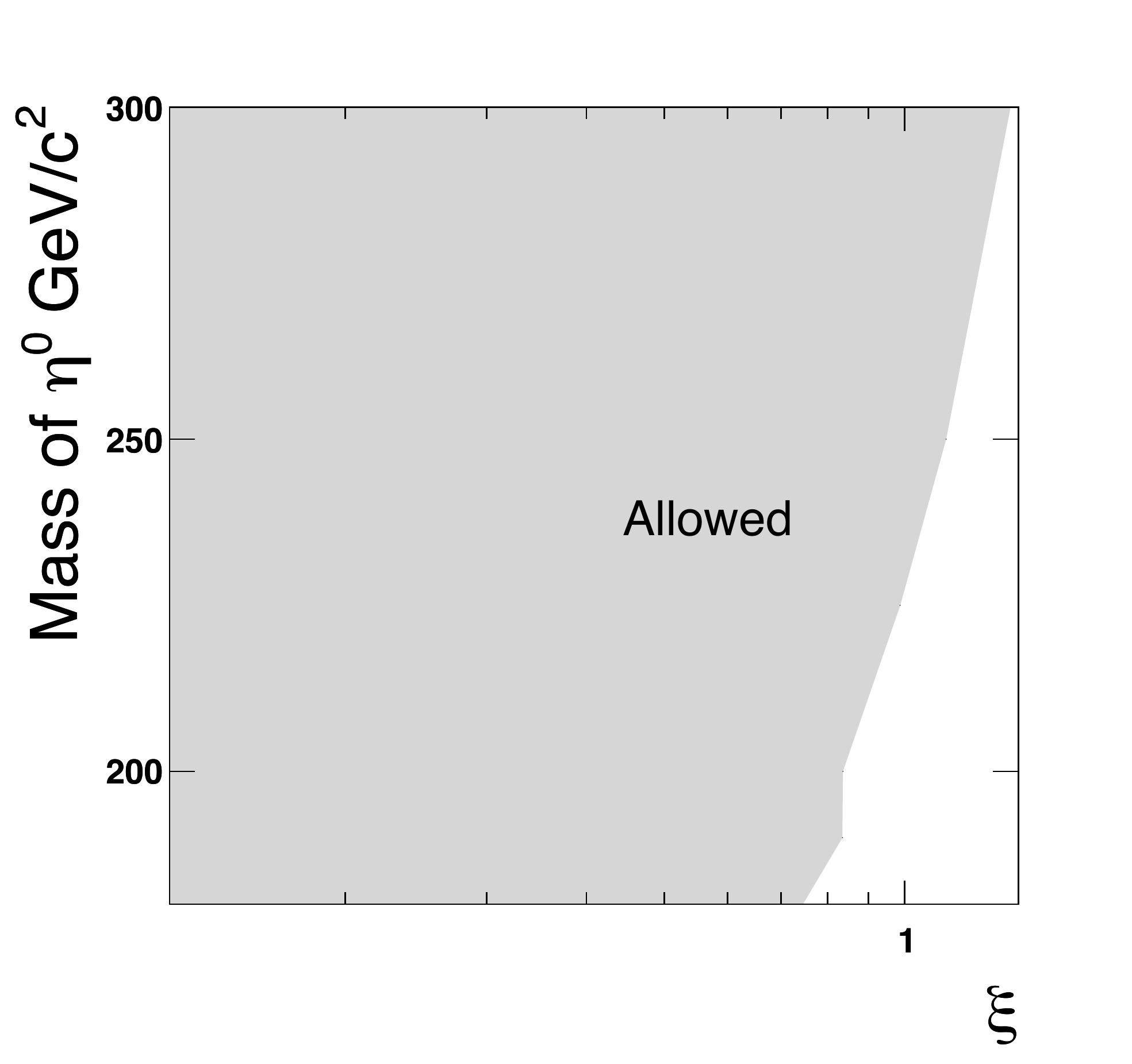}\\
\includegraphics[width=7cm]{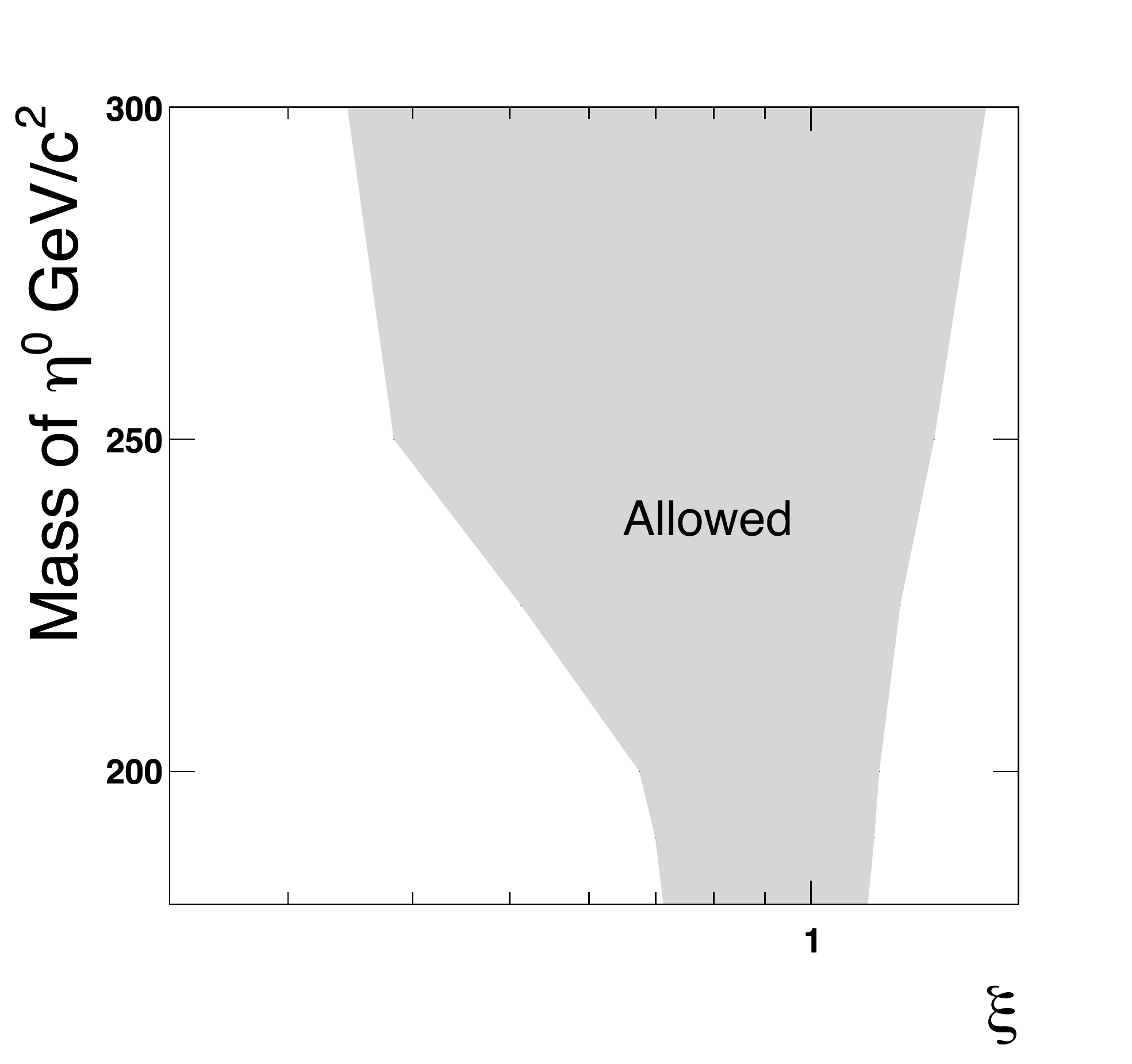}
\caption{Expected 95\% CL allowed regions in the $\xi - m_{\eta^0}$ plane
for 2 fb$^{-1}$  data. Top: for background-only hypothesis using 
$\xi=0$ in the Monte Carlo experiments. Bottom: for
signal-plus-background hypothesis with $\xi=1$ in the Monte Carlo experiments, 
as explained in the text.}
\label{fig:sens}
\end{figure}

Prior to any analysis of the Tevatron data, we can evaluate the expected
sensitivity of the dataset, which indicates the strength of the
measurement or exclusion that either Tevatron experiment could make. Following the
Feldman-Cousins prescription~\cite{feldcous}, we use Monte Carlo
experiments to construct bands which contain 95\% of the fitted
values of $\xi$ at
various true values of $\xi$ for a specific mass of $\eta^0$, see
Fig.~\ref{fig:fc}.
The confidence band in $\xi$ for an individual experiment is the
{\it vertical} band at the fitted $\xi$. For example, a fit value
of $\xi = 1$ would correspond to a 95\% CL band of $\xi \approx 0.7$
to $\xi \approx 1.2$.

The expected sensitivity to $\xi$ is the mean vertical 95\% CL band in
$\xi$ from Monte Carlo experiments. We evaluate the expected sensitivity
for both the background-only (using $\xi=0$ for the Monte Carlo experiments) 
and the signal-plus-background
hypothesis with $\xi=1$, see Fig.~\ref{fig:sens}.
In the case of the background-only hypothesis, the expected allowed
region includes $\xi = 0$, so the result would be interpreted as an
upper limit on $\xi$.
In the case of the signal-plus-background hypothesis, $\xi=0$ is
expected to be excluded at greater than 95\% CL, indicating strong
sensitivity to the presence of $\eta^0 \to tu$ decays if they exist
in the data.

To summarize, we have performed a detailed signal-to-background analysis of
same-sign
leptons signals at the Tevatron, originating from same-sign top quark pair
production
in models of MxFV. We have shown that the current Tevatron dataset has strong
sensitivity to the same-sign lepton signature $l^\pm l^\pm b \missET$,
for natural values of the MxFV coupling, i.e.
$\xi \sim {\cal O}(1)$, and a mass around 200 GeV of the scalar that
mediates MxFV.
In particular, after an event selection corresponding to an acceptance of
0.5\%
we expect 8-16  $l^\pm l^\pm b \missET$ events (from
same-sign top quark pair production) for $m_{\eta^0}\approx 225-180$ GeV/c$^{2}$, respectively, over about 4-5 background events.  As there are no current limits on this MxFV signals,  we urge an analysis of the Tevatron data in this channel.

A similar analysis for the LHC is beyond the scope of this paper.
We can, however, estimate the LHC sensitivity to the same-sign leptons
$l^\pm l^\pm b \missET$ signal. In particular, for the LHC we find
$\sigma(pp \to tt +nj +X) \sim {\cal O}(100)$ pb for $\xi \sim 1$ and
a several hundred GeV $\eta^0$. Thus, based on the present analysis,
after an event
selection with an acceptance around 0.5\% and an early stage integrated luminosity of 
10 fb$^{-1}$, we expect about 5,000 $l^\pm
l^\pm b \missET$ signal events. The background at the
LHC is expected to be dominated by the $t \bar t$ production which has a
cross-section about 1,000 times larger than at the Tevatron. Thus
scaling the Tevatron $t\bar{t}$ background by a factor of 1,000 (see Table II), we expect about 2,000
$l^\pm l^\pm b \missET$ background events. We hope to return to these questions in future work.

Finally, there are many other channels which are also worth analyzing, see \cite{ourMxFV}. 
For example, single top quark production in association with a wrong-sign b jet 
and a light jet, resonance production of $\eta^+$ (i.e., the charged component of $\Phi_{FV}$) and $t\bar{t}$ production which occurs at a similar rate as the same-sign $tt$ production in these models, albeit with a larger background.

{\it Acknowledgments:} AR is supported in part by NSF Grants No.~PHY-0354993 and PHY-0653656. SBS is supported in part by NSF Grants No.~PHY-0653656 (UCI), PHY-0709742 (UCI) and by the Alfred P. Sloan Foundation.


\begin{thebibliography}{99}

\bibitem{ourMxFV} S. Bar-Shalom and A. Rajaraman, hep-ph/0711.3193.

\bibitem{cdfss} A. Abulencia {\it et al.}, Phys. Rev. Lett. 98, 221803 (2007).D. Acosta {\it et al.}, Phys. Rev., Lett. 93, 061802 (2004)

\bibitem{cdfdet}
 A. Abulencia {\it et al.}, J. Phys. G {\bf 34}, 2457 (2007).,
\bibitem{ttbar} D. Acosta {\it et al.} (CDF Collaboration), Phys. Rev. D {\bf 71}, 052003 (2005).
\bibitem{secvtx} A. Abulencia {\it et al}., Phys. Rev. D 74, 072006 (2006), arXiv:hep-ex/0607035.
\bibitem{calchep} A.Pukhov {\it et al.}, Preprint INP MSU 98-41/542.
\bibitem{pythia} T. Sjostrand {\it et al.}, Comput. Phys. Commun. 238 135 (2001).
\bibitem{alpgen} M.L. Mangano, M. Moretti, F. Piccinini, R. Pittau, A. Polosa, JHEP 0307:001,2003, hep-ph/0206293.
\bibitem{feldcous} Feldman and Cousins, Phys. Rev. D 57, 3873 (1998).
\end{thebibliography}
\end{document}